# High field paramagnetic Meissner effect in $Mo_{100-x}Re_x$ alloy superconductors

Shyam Sundar[1,2], M K Chattopadhyay[1,2], L S Sharath Chandra[1] and S B Roy[1,2]

[1] Magnetic and Superconducting Materials Section, Raja Ramanna Center for Advanced Technology, Indore 452 013, India
[2] Homi Bhabha National Institute, Raja Ramanna Center for Advanced Technology, Indore 452 013, India

**Abstract**

We have performed an experimental study on the temperature and field dependence of magnetization of a series of superconducting $Mo_{100-x}Re_x$ alloys. Our studies reveal the presence of high field paramagnetic effect (HFPME) in these low temperature superconductors. The results of our studies indicate that the HFPME in the $Mo_{100-x}Re_x$ alloys is related to the inhomogeneous distribution of strong and weak flux-line pinning centers, and the flux compression resulting due to the same while cooling down the samples in the presence of high magnetic fields. The results are complemented by the studies on the temperature dependence of the electrical resistivity and heat capacity of these alloys in different constant magnetic fields. We compare our findings with the studies reported in literature on both low $T_C$ and high $T_C$ superconductors.

**Introduction**

While a diamagnetic response to applied magnetic field (H) is the well-known signature of superconductivity, a net paramagnetic response has been observed in many superconductors when they are cooled below the normal state to superconducting state transition temperature ($T_C$) in the presence of a magnetic field [1, 2]. The effect is commonly referred to as the 'Paramagnetic Meissner Effect' or PME [1, 2] and has received a lot of attention after it was observed in the high-$T_C$ superconductors [3]. The effect is, however, observed in the conventional superconductors as well [1, 4], and in fact, a paramagnetic response over a certain temperature (T) regime below the $T_C$ was reported in a single crystal of Sn in 1943 [5]. Paramagnetic response was also observed in other elemental and alloy superconductors like Sn, Hg, Nb, Nb–Ta and Pb–Tl in the following couple of decades (1950s–70s) [6, 7]. Literature reports suggest that the phenomenon of PME observed in different superconductors may be classified into two categories. These are the low field PME (LFPME; also called the Wohlleben effect) observed in magnetic fields of the order of 0.1 mT or even less (below the lower critical field HC1), and the high field paramagnetic effect (HFPME) that may be observed in few T (or more) magnetic fields [8, 9].

The LFPME in the high-$T_C$ superconductors has been explained in terms of the generation of spontaneous supercurrents due to the combined effects of randomly distributed Josephson π-junctions between the superconducting grains, and the d-wave pairing symmetry [2, 10, 11]. The same explanation, however, does not hold for the qualitatively similar LFPME observed in the Nb discs and thin films where the effect is reported to be related to the sample geometry and surface related effects [4, 12–14]. Similar reports on the sample geometry and surface related LFPME are found in some of the high-$T_C$ superconductors as well [15]. The LFPME in such cases are explained in terms of the compression of magnetic flux trapped inside the superconductor. This may occur as a result of inhomogeneous cooling (in the presence of magnetic field) so that the sample-edges become superconducting before the main bulk, and then the expulsion of flux from the edges causes flux-trapping inside the sample. Upon further cooling, this flux-free edge-region is broadened which causes compression of flux [16]. Flux-trapping and the subsequent flux-compression may also be possible due to strong flux-pinning sites at the sample surface [4, 12], the formation of a giant vortex state at the sample surface (single giant vortex or multiple vortices developing from the microstructural defects at the surface) [4, 17], or due to flux capture resulting from special geometry effects [18]. While the formation of the giant vortex state has been found to explain the LFPME of Al-discs [19], the LFPME observed in the Josephson junction arrays fabricated from Nb and $Al_2O_3$ [20, 21] was explained with the help of a magnetic-screeningmodel derived for multiply-connected superconductors. This model accounted for LFPME without invoking d-wave superconductivity. The LFPME observed in the thin films of Nb and YBCO may be explained on the basis of this concept [14]. Apart from the more intrinsic causes mentioned above, LFPME observed in the $Nd_{2-x}Ce_xCuO_y$ superconductors are thought to originate due to flux-trapping and compression resulting from an inhomogeneous distribution of superconducting and non-superconducting regions [22]. An extrinsic LFPME has also been generated artificially in the multi-phase Sn–In samples having different $T_C$ values at different locations [23]. Recently, the LFPME observed in the $La_{1.85}Sr_{0.15}CuO_4$ single crystals have been related to the anisotropy of upper critical field and the presence of a surface superconducting state [24].

The HFPME was first observed in the high-$T_C$ superconductors by Rykov et al [25], and in contrast with the LFPME this former effect is observed in the mixed state ($H_{C1} < H < H_{C2}$) of the concerned superconductors [8, 9, 13, 25–28]. Similar to the case of LFPME, the values of magnetization (M) measured while cooling the sample in the presence of a magnetic field is lower than that measured while warming up in the same field in the superconducting state of the samples exhibiting HFPME. The other characteristic features of HFPME are the dependence on the cooling rate and the size of the samples, suggesting that the paramagnetic response might be induced by flux compression, which could originate from inhomogeneous cooling and/or inhomogeneous distribution of strong and weak flux-line pinning centres coupled with flux creep [8, 9, 25]. HFPME has mostly been reported in the $YBa_2Cu_3O_{7-\delta}$ [8, 9, 25] and $MgB_2$ samples [26, 27], and these superconductors have been found to exhibit both HFPME and LFPME [9, 27, 29]. While the HFPME in the $MgB_2$ superconductors is believed to be related to the anisotropy in the $H_{C2}$ and the random orientation of the grains [27], the LFPME in these superconductors is thought to have an intrinsic origin [29]. The HFPME is, however, rarely observed in the conventional low $T_C$ superconductors. Among the low $T_C$ superconductors, the thin films of Nb are known to exhibit PME up to 0.12 T [13], and we have recently observed HFPME in bulk Ti–V alloy superconductors in the field range of 0.5–1.5 T [30]. In the Ti–V alloys, the HFPME appears below a characteristic temperature that shifts towards a lower temperature with increasing applied magnetic field, and the magnetization measured at a constant temperature after cooling down the sample in the presence of magnetic field increases with time in the HFPME regime. We found the effect related to the inhomogeneous distribution of the stress induced martensitic phase formed in these alloys during cold-work. The HFPME disappears when this stress induced phase is removed by annealing [30]. Therefore, the HFPME in the Ti–V alloys have a somewhat extrinsic origin. One now needs to investigate whether a relatively intrinsic phenomenom such as the inhomogeneous distribution of strong and weak flux-line pinning centres can cause HFPME in the low $T_C$ materials, or is the phenomenon of completely extrinsic origin in these materials. To address this question we now investigate into another system of alloys, namely the $Mo_{100-x}Re_x$ alloys, and look for the signatures of HFPME in this system.

In the present work, we study the temperature and field dependence of magnetization of a series of superconducting $Mo_{100-x}Re_x$ alloys, and also complement these studies with the measurement of the temperature dependence of electrical resistivity ($\rho$) and heat capacity (C) of these alloys in different constant magnetic fields. We identify the characteristic signatures of HFPME in these alloys, and explain our results in terms of the inhomogeneous distribution of strong and weak flux-line pinning centres, and the flux-compression resulting due to the same while cooling down the samples in the presence of high magnetic fields.

## Experimental

The $Mo_{75}Re_{25}$, $Mo_{60}Re_{40}$, $Mo_{52}Re_{48}$ alloys were synthesized by melting high purity (99.95+%) Mo and Re (ACI Alloys, USA) taken in atomic proportions. The melting was performed under high purity Ar atmosphere in an arc-melting furnace. The samples were flipped and re-melted six times to ensure the homogeneity. The loss of mass during the melting procedure was less than 0.1%. The prepared samples were cut in required shapes and sizes with the help of a diamond wheel cutter and a spark wire cutter for different experimental measurements. While most of the measurements were performed on the as-cast samples, the $Mo_{60}Re_{40}$ alloy was measured both in the as-cast and annealed condition. The annealing was done (in sealed quartz ampoule, with the sample wrapped in Ta foil) at 1250 °C for 20 h in Ar atmosphere, which was followed by slow-cooling down to room temperature [31]. The structural characterization of the present $Mo_{100-x}Re_x$ alloy samples was done with the help of x-ray Diffraction (XRD), optical metallography, and energy dispersive x-ray (EDX) measurements. These XRD experiments were performed in a standard diffractometer (Rigaku Corporation, Japan: Geigerflex model) using Cu K$\alpha$ ($\lambda$=1.5418 Å) radiation. For optical metallography, small pieces of samples were mounted on moulds prepared from epoxy (Resin+Hardener) and then ground with the help of silicon carbide abrasive papers. Properly grounded moulds were then polished to 0.5 μm roughness with the help of diamond paste. Polishing of the sample was done by holding the moulds on rotating Nylon cloth. An etchant solution of H2O2 (50%) and NH4OH (50%) were used for the visualization of the microstructure of the polished samples. The metallographical characterization of the samples was done using a high power optical microscope (Leica, DMI 5000M). The chemical compositions of the samples were experimentally confirmed through EDX analysis performed using a Philips XL-30 pc machine. Within the experimental resolution of the EDX measurements, the chemical compositions of the alloys were not found to vary over the bulk of the sample.

The magnetization (M) measurements as functions of temperature (T) and magnetic field (H) were performed using a Vibrating Sample Magnetometer (VSM; Quantum Design, USA). The M versus T measurements were performed following the zero field cooled (ZFC) warming, field cooled cooling (FCC) and field cooled warming (FCW) protocols. In the ZFC protocol, the sample was cooled in zero field from 20 K down to the lowest temperature of measurement (2 K). The field was applied at 2 K and the M versus T measurements were performed in this field while warming up the sample up to 20 K. After the ZFC warming measurements up to 20 K, the sample was cooled down to 2 K in the same magnetic field, and the M versus T measurements were continued during this cool-down process to obtain the FCC curve. After the FCC experiments, the M versus T measurements were further continued in the same magnetic field while warming up the sample back to 20 K. This last protocol of measurements is called the FCW. The isothermal M versus H measurements were performed while varying µ0H sequentially from 0 to 7 T, from 7 T to -7 T, and then from -7 T to 7 T, starting from an initial ZFC state.

The T dependence of C in different H values for the $Mo_{100-x}Re_x$ alloys were measured in a Physical Property Measurement System (PPMS; Quantum Design, USA) using a two-τ relaxation technique. The T dependence of ρ for different H values were measured using standard four-probe configuration in different standard cryostats (the PPMS, a cryostat from the Oxford Instruments, UK, and another cryostat from the American Magnetics, USA). The T dependence of both ρ and C were measured while cooling down the samples below 20 K (in different fields).

## Results and discussion

Structural and metallographic characterization

The XRD patterns for the as-cast samples of $Mo_{75}Re_{25}$ and $Mo_{60}Re_{40}$ alloys indicate that these alloys are single-phase, and have formed in the body centred cubic (bcc) β-phase structure. These alloys have been used in our previous study where the XRD patterns have been presented [32]. On the other hand, the XRD patterns of the annealed $Mo_{60}Re_{40}$ alloy and the as-cast $Mo_{52}Re_{48}$ alloy indicate the presence of the β, σ (tetragonal), and χ (complex bcc) phases in these samples. The annealed $Mo_{60}Re_{40}$ alloy contains about 75% β phase, 20% σ phase and 5% χ phase. The as-cast $Mo_{52}Re_{48}$ alloy, on the other hand contains approximately 50, 40 and 10% of β, σ and χ phases respectively.

We have performed optical metallography studies on different portions of each of the present samples. Only a few representative micrographs are presented in figure 1 for the sake of conciseness. Figure 1(a) shows an optical micrograph for the as-cast $Mo_{75}Re_{25}$ alloy (single phase). Large grains with varying sizes and a network of dislocations [33–35] are observed in this micrograph. Figure 1(b) shows an optical micrograph for the as-cast $Mo_{60}Re_{40}$ alloy (single phase), which also shows large grains with varying sizes and a network of dislocations similar to the as-cast $Mo_{75}Re_{25}$ alloy. This sample was over-etched to reveal the signature of the dislocations. Our optical metallography studies show that while the grain sizes in the as-cast $Mo_{75}Re_{25}$ and $Mo_{60}Re_{40}$ alloys (single-phase) mostly vary from 100–200 µm to 500–600 µm, some of the very large grains have dimensions of about 2000 µm. Thus the density of the grain boundaries varies strongly over the sample-surface for these alloys.

Comparatively, the spatial distribution of the dislocations for these alloys seems to be more uniform. Figure 1(c) shows an optical micrograph for the annealed $Mo_{60}Re_{40}$ alloy, which contains β, σ and χ phases. The average grain size in the annealed $Mo_{60}Re_{40}$ alloy is 50–150 μm, which is much smaller as compared to that of the single-phase alloys. Apart from the grain boundaries visible in figure 1(c), it appears that the σ and χ phases have been precipitated around the dislocation networks in this alloy. The precipitation of the σ phase along the dislocation networks in the $Mo_{100-x}Re_x$ alloys have also been reported earlier [35]. Figure 1(d) shows an optical micrograph for the as cast $Mo_{52}Re_{48}$ alloy (containing β, σ and χ phases), which shows a eutectoid structure consisting of a matrix, a fine lamellar pattern initiating from the grain boundaries of the matrix, and a nodular pattern (marked by rings in figure 1(d) [36]. We expect the matrix to be the β phase, which is the original phase in this eutectoid composition, since the other eutectoid-microstructures are expected to start from the grain boundaries of this original phase and spread inside the samples [36]. As the fine lamellar pattern is more abundant in figure 1(d) as compared to the nodular pattern, and since the XRD pattern for the as-cast $Mo_{52}Re_{48}$ alloy indicates that the σ phase has much higher volume fraction as compared to the χ phase, we believe that the fine lamellar pattern is due to the σ phase in the alloy. The relatively rare nodular pattern is thus ascribed to the χ phase. The average grain size in the (multi-phase) $Mo_{52}Re_{48}$ as-cast alloy (figure 1(d)) is found to be (20–100 μm), which is the smaller than the annealed $Mo_{60}Re_{40}$ alloy. No indication of dislocation network is visible in the optical micrograph of this alloy.

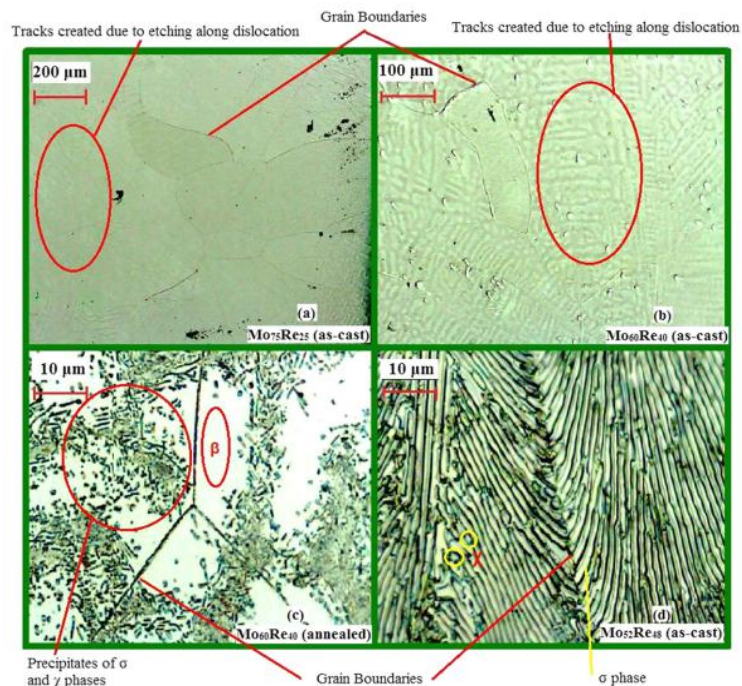

Figure 1 Micrographs showing the results of optical metallography experiments performed on the $Mo_{100-x}Re_x$ alloys.

Temperature and field dependence of magnetization and the signature of HFPME

Figure 2(a) shows the variation of M as a function of normalized T for the as-cast $Mo_{60}Re_{40}$ alloy measured in the ZFC, FCC, and FCW protocols in 10 mT magnetic field, where T is normalized with respect to the superconducting transition temperature $T_C$. The value of $T_C$ for the alloy was earlier obtained by finding the temperature at which the M(T) curves (measured in 10 mT magnetic field) start to drop downwards towards a negative value when the temperature is decreased. For the as-cast $Mo_{60}Re_{40}$ alloy, this value comes out to be 12.78 K. Following the same technique the $T_C$ values for the as-cast $Mo_{75}Re_{25}$, annealed $Mo_{60}Re_{40}$, and ascast $Mo_{52}Re_{48}$ alloys are found to be 9.85, 12.44, and 12.68 K respectively. The $T_C$ values for the present alloys are consistent with the literature reports [37]. The M versus normalized T curves in figure 2(a) shows the features commonly expected for a type-II superconductor [38]. The values of M are negative below the $T_C$, and are smaller (in magnitude) for the FCC and FCW curve as compared to the ZFC curve. Moreover, the FCC curve lies above the FCW curve near the transition region. Very similar features are observed in the M versus normalized T curves for the as-cast $Mo_{60}Re_{40}$ alloy in 0.1 T magnetic field as well. However, the M(T) curves undergo qualitative change as the measuring field is increased to 0.3 T (not shown here). In this field, the FCW curve lies above the FCC curve at T below a cross-over temperature $T_{CR}$. This feature of the FCW curve lying above the FCC curve is also observed in higher fields in the as-cast $Mo_{60}Re_{40}$ alloy, as is shown in figure 2(b) and in the inset to this figure for $\mu_0H = 1$ T. The value of $T_{CR}$ decreases with increasing applied field from 4.47 K for $\mu_0H = 0.3$ T to 2.45 K for $\mu_0H = 1$ T. We believe that the phenomenon continues to exist in still higher fields, but at temperatures below the scope of measurement in our experimental set-up. The inset to figure 2(b) also reveals that when the measurement-protocol is switched from FCC to FCW at 2 K, the measured M shows strong relaxation effects (the FCC M-value below $T_{CR}$ increases nearly linearly with time at constant temperature and field; these curves are not presented here for the sake of conciseness). As has been pointed out earlier, the FCW M(T) curve lying above the FCC curve and the field cooled M showing relaxation effect (increasing M) are characteristic features of HFPME [8, 9, 13, 25–28, 30]. Such relaxation effect is not observed in the inset to figure 2(a) where the applied field is 10 mT. The as-cast $Mo_{75}Re_{25}$ alloy is also found to exhibit HFPME, and a value of $T_{CR} = 2.2$ K was obtained for this alloy for $\mu_0H = 0.5$ T. The effect is not observed in lower fields (viz. 0.1 T) in this alloy. In higher fields ($\mu_0H = 0.7$ T, say), the value of $T_{CR}$ for the as-cast $Mo_{75}Re_{25}$ alloy goes below the range of temperature which can be measured in our experimental set-up. However, the field cooled M for $\mu_0H = 0.7$ T does show relaxation effect (increasing M) at 2 K, indicating the presence of HFPME in this field in the as-cast $Mo_{75}Re_{25}$ alloy. The M versus normalized T curves for the as-cast $Mo_{75}Re_{25}$ alloy, including the signatures of HFPME look qualitatively very similar to those for the as-cast $Mo_{60}Re_{40}$ alloy. We are, therefore, not presenting the M versus normalized T curves for the as-cast $Mo_{75}Re_{25}$ alloy separately. In figures 2(c) and (d) we present the M versus normalized T curves for the annealed $Mo_{60}Re_{40}$ alloy and the as-cast $Mo_{52}Re_{48}$ alloy for $\mu_0H = 1$ T. While there is no signature of PME for $\mu_0H = 10$ mT for these alloys (data exists, but not shown here for the sake of conciseness), figures 2(c) and (d) and their insets show the same signatures of HFPME as mentioned above. Though the relaxation effect (increase) in the field cooled M when the measurement-protocol is switched from FCC to FCW at 2 K is not clearly visible in the inset to figure 2(c), this effect is actually present in the data and is observable at a higher magnification. While the characteristic features of HFPME are observed in all the four alloys presented here, comparison of the insets to the figures 2(b)–(d) reveals that the M(T) behaviour for the single-phase and multiphase $Mo_{100-x}Re_x$ alloys below the $T_{CR}$ are not exactly similar. Firstly, in the single phase alloys, the $T_{CR}$ is observed at much lower normalized T as compared to the multiphase alloys. Secondly, the FCW M-values below the $T_{CR}$ of the single-phase $Mo_{100-x}Re_x$ alloys do not vary appreciably with increasing T. This is clearly observed in the inset to figure 2(b). In this inset, the variation of M along the FCW curve in the regime from the lowest measured temperature to the $T_{CR}$ (the temperature at which the FCC and FCW curves cross each other) is nearly same as the width of/ jitter in the experimental curve. This is not true for the case of the multiphase alloys (see the insets to the figures 2(c) and (d)) where the FCW M-values below the $T_{CR}$ (temperature at the right edge of these insets approximately) increase appreciably with increasing temperature.

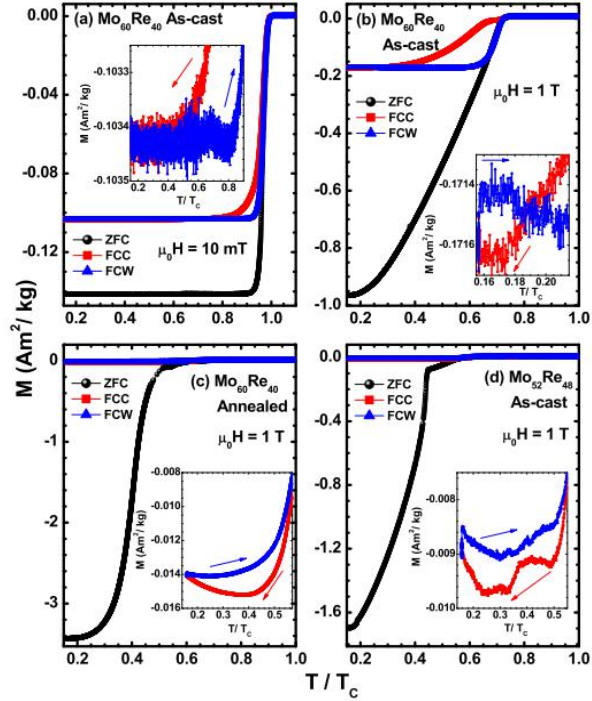

Figure2: Variation of magnetization of the $Mo_{100-x}Re_x$ alloys as a function of normalized temperature, showing the signature of HFPME in 1 T magnetic field and the absence of the same in 10 mT field.

While the value of $T_{CR}$ above 1 T magnetic field for the present single-phase $Mo_{100-x}Re_x$ alloys goes below the range of temperature which can be measured in our experimental set-up, the $T_{CR}$ can be measured up to much higher fields in the present multiphase $Mo_{100-x}Re_x$ alloys. Figure 3 shows the M versus normalized T curves for the annealed $Mo_{60}Re_{40}$ alloy and the as-cast $Mo_{52}Re_{48}$ alloy for $\mu_0 H = 2$ T. In this field, both the alloys exhibit the characteristic features of HFPME discussed above. Moreover, for both these alloys, the values of M for the FCC and FCW curves are positive (in the HFPME regime), though the alloys are still in the superconducting state and the value of M is negative for the major portions of the ZFC M(T) curves. In the annealed $Mo_{60}Re_{40}$ alloy, the HFPME may be observed over a large magnetic field regime starting from 0.3 T to beyond 5 T. Within the HFPME regime (below the $T_{CR}$) in this alloy, the temperature gradient of the FCW M versus normalized T curves become more and more positive with increasing magnetic field. This effect may be understood by comparing figures 2(c) and 3(a). The M versus normalized T curves for this alloy for all the other measured fields are not shown here for the sake of conciseness. For the as-cast $Mo_{52}Re_{48}$ alloy, on the other hand, the temperature gradient of the FCW M versus normalized T curve below the $T_{CR}$ becomes clearly negative as the field is increased from 1 T to 2 T (see figures 2(d) and 3(b)). We have already mentioned that the FCW M versus normalized T curve does not exhibit an appreciable temperature dependence below the $T_{CR}$ in the case of the single phase $Mo_{100-x}Re_x$ alloys (as-cast $Mo_{75}Re_{25}$ and $Mo_{60}Re_{40}$ alloys). Our results indicate that this situation does not change appreciably with increasing magnetic field in the case of the single phase $Mo_{100-x}Re_x$ alloys. Thus the FCW M(T) behaviour of the single-phase and multiphase $Mo_{100-x}Re_x$ alloys are qualitatively different within the HFPME regime. Even for the multiphase alloys, the annealed $Mo_{60}Re_{40}$ alloy behaves differently from the as-cast $Mo_{52}Re_{48}$ alloy within the HFPME regime. While we will subsequently discuss more on these differences in the FCW M(T) behaviour of different $Mo_{100-x}Re_x$ alloys, we would like to rule out at this stage the possibility that the positive values of M in the HFPME regime (in 2 T magnetic field), and the negative temperature gradient of the M(T) curves for the as-cast $Mo_{52}Re_{48}$ alloy are due to the presence of magnetic impurities in the our $Mo_{100-x}Re_x$ alloys. This was tested by measuring the isothermal field dependence of M for all the present alloys at temperatures just above the $T_C$ (13–15 K). These M(H) curves are completely linear, indicating the absence of magnetic impurities in the above alloys. We now investigate into the M(H) behaviour of the present alloys below their $T_C$ and try to correlate their characteristic features with the HFPME described above.

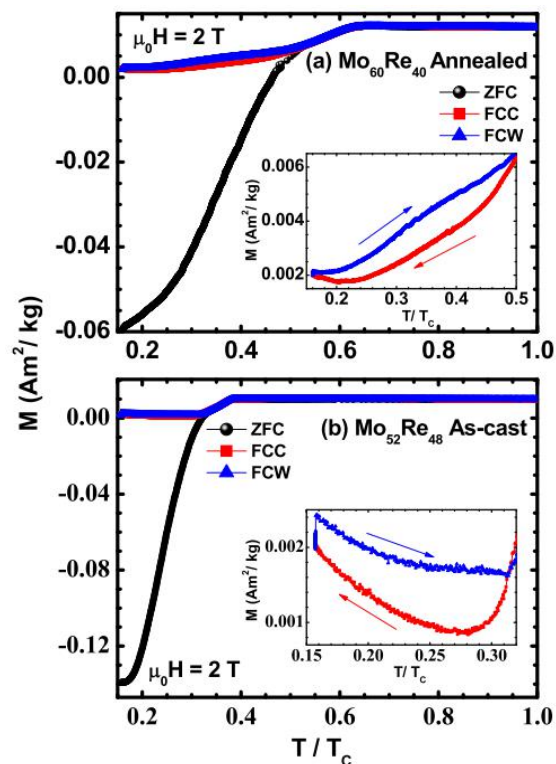

Figure3 Variation of magnetization of the multiphase $Mo_{100-x}Re_x$ alloys as a function of normalized temperature, showing the signature of HFPME in 2 T magnetic field.

Figure 4(a) shows the isothermal M(H) curves for the ascast $Mo_{60}Re_{40}$ alloy at 2 K. A large magnetic hysteresis is observed in this figure, which reduces quite rapidly with increasing magnetic field. The high-field side of this M(H) curves are shown with larger magnification in figure 4(b). This figure shows that though the major magnetic hysteresis nearly disappears by $\mu_0H = 2.74$ T, a small amount of irreversibility between increasing and decreasing magnetic fields persists up to much higher magnetic fields (close to 5 T). Figure 4(b) also reveals that an indication of $H_{C2}$ in the form of a distinct deviation [39] from the M(H) behaviour of the normal state (in the high H-regime) is not clearly observed in the as-cast $Mo_{60}Re_{40}$ alloy. The deviation is rounded-off over a large H-regime, and qualitatively similar features were observed for the as-cast $Mo_{75}Re_{25}$ alloy as well (not shown here for the sake of conciseness). Figure 4(c) shows the isothermal M(H) curves for the as-cast $Mo_{52}Re_{48}$ alloy at 2 K, and figure 4(d) shows the high-field side of the same curves with larger magnification. In this case, however, a distinct deviation [39] from the M(H) behaviour of the normal state (in the high H-regime) is clearly visible. The major part of the magnetic hysteresis (figure 4(c)) finishes off at a lower field than this point of deviation, while a small amount of tailing magnetic hysteresis persists up to much higher fields (figure 4(d)). The M(H) curves for the annealed $Mo_{60}Re_{40}$ alloy also show features qualitatively similar to the as-cast $Mo_{52}Re_{48}$ alloy, but are not shown here for the sake of conciseness. Evidently, the value of $H_{C2}$ for the present series of alloys cannot be obtained from the M(H) curves in a straight-forward and consistent manner. On the other hand, the tailing magnetic irreversibility persisting up to fields presumably higher than the $H_{C2}$ may be due to the presence of the third critical field HC3 (surface superconductivity) reported in the $Mo_{100-x}Re_x$ alloys [31]. In order to obtain a clear picture on this issue, we have estimated the normal to superconducting transition temperatures from the C(T) and ρ (T) curves for the present alloys obtained in different constant magnetic fields. These results are discussed below.

Figure 5 shows the ρ(T) and C(T) curves for the as-cast $Mo_{60}Re_{40}$ alloy in 1.5 T magnetic field. The normal state to superconducting state phase transition in these alloys shows up while cooling down as a sharp drop in the ρ(T) curve and a rise (towards a peak) in the C(T) curve. The transition temperature is obtained from the maximum/minimum observed in the temperature derivative of the ρ(T) and C(T) curves. The insets to the figures 5(a) and (b) show the temperature derivatives of the ρ(T) and C(T) curves for the as-cast $Mo_{60}Re_{40}$ alloy in 1.5 T magnetic field. These insets highlight the difference between the T values corresponding to the maximum/minimum observed in the temperature derivatives of the ρ(T) and C(T) curves. Experimental curves similar to those presented in figure 5 have been obtained for all the four present alloys in different magnetic fields. The T and H values corresponding to the maximum/minimum observed in the temperature derivatives of the ρ(T) and C(T) curves are used to plot the T-dependence of $H_{C2}$ for the respective alloy, which is shown in figure 6.

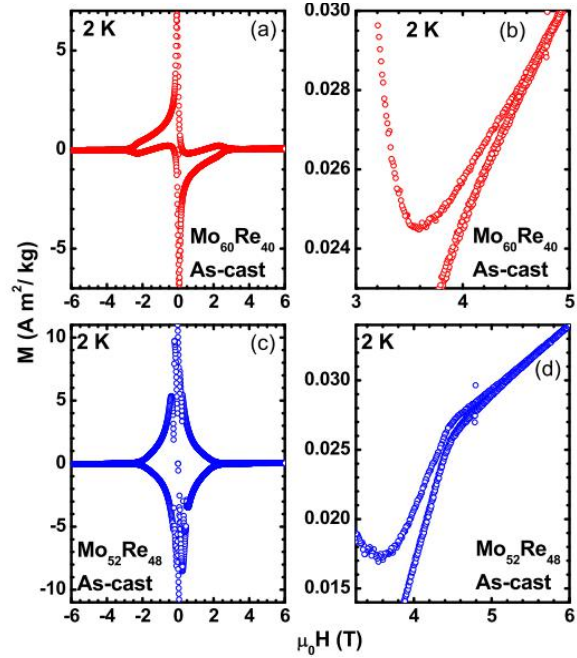

Figure 4 Isothermal field dependence of magnetization of the as-cast $Mo_{60}Re_{40}$ and $Mo_{52}Re_{48}$ alloys at 2 K. The panels (a) and (c) show the major part of the magnetic irreversibility. The panels (c) and (d) show the corresponding high-field regimes in magnified scales.

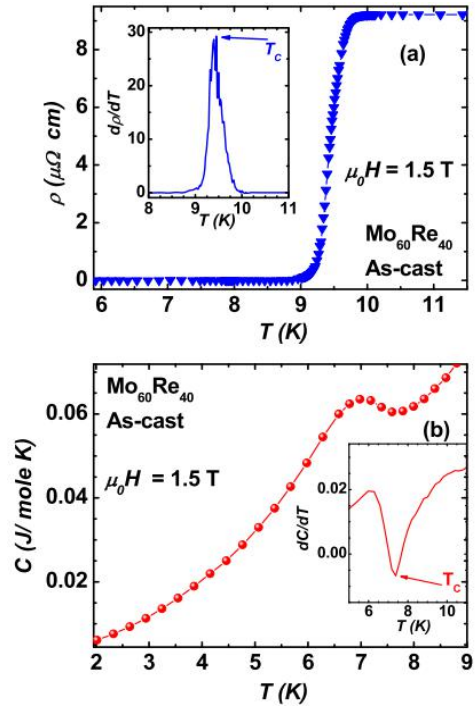

Figure 5 Temperature dependence of resistivity ρ(T) and heat capacity C(T) for the as-cast $Mo_{60}Re_{40}$ alloy in 1.5 T magnetic field. The insets show the temperature derivatives of these ρ(T) and C(T) curves.

In figure 6, we present the H–T phase diagrams for the present $Mo_{100-x}Re_x$ alloys in order to correlate the HFPME with the superconducting properties. These phase diagrams show a comparison between the field dependence of $T_{CR}$ and the T-dependence of $H_{C2}$ obtained from the C(T) and ρ(T) curves for the present alloys (as described above). It is observed in figures 6(a), (c) and (d) that except in the low field regime, the $H_{C2}(T)$ line obtained from the ρ(T) curves lies well above that obtained from the C(T) curves for the ascast $Mo_{75}Re_{25}$ and $Mo_{60}Re_{40}$ alloys, and the as-cast $Mo_{52}Re_{48}$ alloy. Only for the annealed $Mo_{60}Re_{40}$ alloy, these two $H_{C2}(T)$ lines match with each other. It is well known that for the ρ(T) measurements, the (four) probes are on the sample surface, and this measurement may be affected by surface-related phenomena. The C(T) measurement technique (see the experimental section), on the other hand, measures only the bulk phenomena. Therefore we infer that surface superconductivity may be present in the as-cast $Mo_{75}Re_{25}$ and $Mo_{60}Re_{40}$ alloys, and the as-cast $Mo_{52}Re_{48}$ alloy, but not in the annealed $Mo_{60}Re_{40}$ alloy. It is to be noted in this context that for the present alloys, if the $H_{C2}$ is defined as the field where the long tailing hysteresis in the isothermal M(H) curves disappears, then this value of $H_{C2}$ nearly matches with that obtained from the ρ(T) measurements. Figure 6 also makes it clear that the HFPME in the present $Mo_{100-x}Re_x$ alloys does not originate from surface superconductivity. The experimental signatures for HFPME are very robust for the annealed $Mo_{60}Re_{40}$ alloy, and this alloy does not indicate the presence of surface superconductivity. In figure 6, we have not plotted the HC1(T) line for the present alloys. The HC1 values for these alloys were estimated from the isothermal M (H) curves, following a standard method described in our previous work [32]. These values were less than 70 mT at 2 K for all the present alloys, and they are not really visible in the same scale as that of the $H_{C2}(T)$ lines. Thus figure 6 also confirms that the present HFPME is observed well inside the mixed state of the superconductors, which indicates that the phenomenon is interlinked with flux-line pinning.

The present magnetization measurements were performed on samples of small dimensions (in the range of 0.5–1.5 mm), in the presence of He exchange gas, in a standard experimental set-up, using standard protocols, and slow cooling rate (0.1 K min-1 for the M(T) measurements). Any appreciable inhomogeneity in cooling these (metallic) samples is not really expected in the present measurements. We therefore need to look into more intrinsic physical phenomena that could lead to the HFPME in the present alloys. We have recently analyzed the field dependencies of the critical current density (JC) and pinning force density of a wide range of $Mo_{100-x}Re_x$ alloy compositions [40]. In this study we found that the annealed $Mo_{60}Re_{40}$ and the as-cast $Mo_{52}Re_{48}$ alloys (multiphase) have appreciably higher JC and higher pinning force density as compared to the (single-phase) as-cast $Mo_{75}Re_{25}$ and $Mo_{60}Re_{40}$ alloys [40]. In figure 7 we present for example the field dependence of pinning force density {FP(H), estimated from isothermal M(H) curves [40]} for the as-cast $Mo_{75}Re_{25}$ and $Mo_{52}Re_{48}$ alloys obtained at 2 and 4 K. The difference in the pinning force densities in the two alloys is clearly visible in this figure. In figure 6, on the other hand, we see that the HFPME may be observed in a higher H–T regime in the annealed $Mo_{60}Re_{40}$ and the as-cast $Mo_{52}Re_{48}$ alloys as compared with to the as-cast $Mo_{75}Re_{25}$ and $Mo_{60}Re_{40}$ alloys. This indicates that high pinning force density has a strong linkage with the observation of HFPME in the present $Mo_{100-x}Re_x$ alloys.

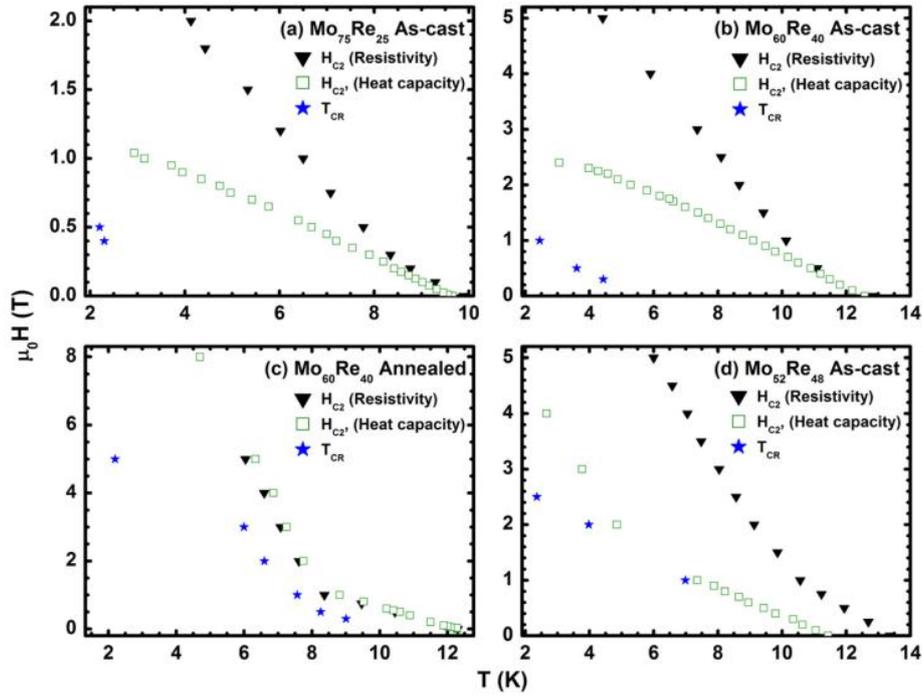

Figure 6 Phase diagrams for the $Mo_{100-x}Re_x$ superconductors showing the characteristic fields and temperatures related to the superconducting properties and HFPME.

In our recent work on the pinning force density of the $Mo_{100-x}Re_x$ alloys, we have found that the flux-line pinning mechanisms in the single-phase $Mo_{100-x}Re_x$ alloys (e.g., the as-cast $Mo_{75}Re_{25}$ and $Mo_{60}Re_{40}$ alloys) is different from the multiphase alloys (e.g., the annealed $Mo_{60}Re_{40}$ and the ascast $Mo_{52}Re_{48}$ alloys). In figure 7 we can see that the FP(H) curve for the as-cast $Mo_{75}Re_{25}$ alloy (single-phase) has a twopeak structure, while that for the as-cast $Mo_{52}Re_{48}$ alloy (multiphase) has essentially a single-peak structure (ignoring some fluctuations coming due to flux jump effects [40]). We have found that this distinction between the single and multiphase $Mo_{100-x}Re_x$ alloys is true for a wide range of alloy compositions [40]. Using standard theories/models [41, 42] we have shown that the dislocation networks are the main flux-line pinning centres in the higher magnetic fields in the case of the single-phase $Mo_{100-x}Re_x$ alloys (the as-cast $Mo_{75}Re_{25}$ and $Mo_{60}Re_{40}$ alloys in the present case). This flux-line pinning mechanism contributes to the major portions of the FP(H) curves, including the high-field peak. In the case of the single-phase $Mo_{100-x}Re_x$ alloys, the grain boundaries are effective in pinning the flux-lines only in very low fields (contributing the low-field peak) [40]. It appears that due to the low grain boundary density in these alloys, the grain boundaries cannot function as major flux-line pinning centres in the higher field regions [40]. Therefore during an FCC experiment there will be large flux-compression [8, 9, 25–27] mostly in the dislocation-loop regions, and flux-depleted regions will be created in the other parts of the samples where the density of the flux-lines will be less than what is expected in an equilibrium-condition. This will open-up the scope for accommodation of extra flux-lines in the samples, and during the FCW measurements the presence of these extra flux lines will suppress the diamagnetic response expected in these superconductors. This will give rise to the observed HFPME. In our previous work we have shown that the single-peak FP(H) curves for the annealed $Mo_{60}Re_{40}$ and the as-cast $Mo_{52}Re_{48}$ alloys (see figure 7) indicate that the grain boundaries are the main flux-line pinning centres in these alloys [40]. While no signature of the presence of the dislocation-loops is observed in the as-cast $Mo_{52}Re_{48}$ alloy, the precipitation of the σ and χ phases around the dislocation networks [35, 40] in the annealed $Mo_{60}Re_{40}$ alloy renders the dislocation networks ineffective in pinning the flux-lines [40]. As such, it is known from reported literature that the σ and χ phases are not very effective in pinning the flux-lines in the $Mo_{100-x}Re_x$ alloys [35, 40]. We believe that the grain boundaries act as the sites for flux-compression in the case of the annealed $Mo_{60}Re_{40}$ and the as-cast $Mo_{52}Re_{48}$ samples. The phenomenon of flux-compression becomes more effective when there is stronger inhomogeneity in the distribution of the strong and weak pinning centres. We have observed in our optical micrographs that the density of the grain boundaries in the $Mo_{100-x}Re_x$ alloys is rather low because of the large grain-size. This gives rise to an inhomogeneous spatial distribution of the stronger pinning centres in the annealed $Mo_{60}Re_{40}$ and the as-cast $Mo_{52}Re_{48}$ alloys. In the as-cast $Mo_{75}Re_{25}$ and $Mo_{60}Re_{40}$ alloys (single phase), on the other hand, the spatial density/distribution of the major pinning centres (dislocation network) effective in the concerned high field regime (where HFPME is observed) is comparatively less inhomogeneous (see figure 1). Because of the rather inhomogeneous spatial density/distribution of the grain boundaries, the HFPME in the annealed $Mo_{60}Re_{40}$ and the ascast $Mo_{52}Re_{48}$ alloys (multiphase) is comparatively more robust against H and T (see figure 2; in 1 T magnetic field, HFPME is observed at $0.6T_C$ and below in the annealed $Mo_{60}Re_{40}$ and the as-cast $Mo_{52}Re_{48}$ alloys, and only below $0.2T_C$ in the as-cast $Mo_{60}Re_{40}$ alloy).

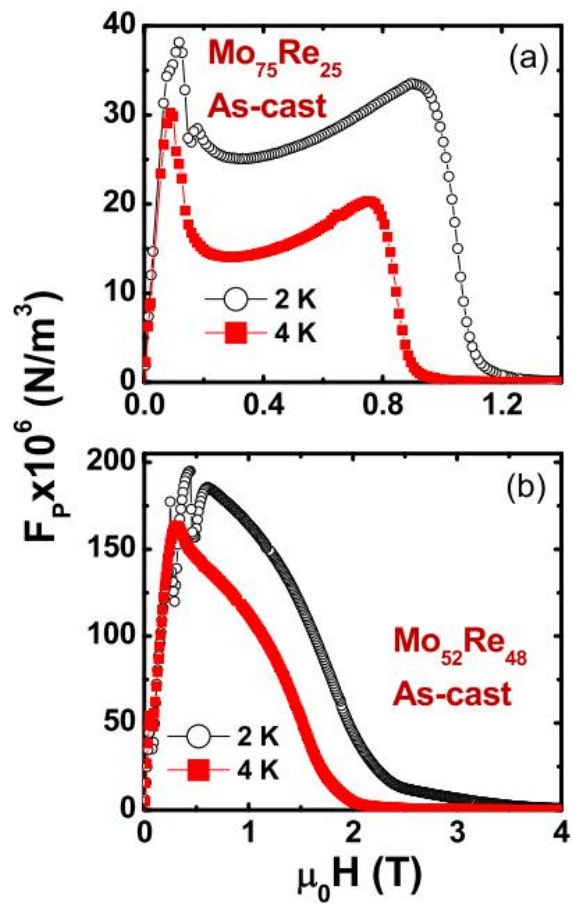

Figure 7 Field dependence of pinning force density of the $Mo_{100-x}Re_x$ alloys.

In our previous study [40] we have seen that the as-cast $Mo_{52}Re_{48}$ alloy has the highest pinning force density among all our $Mo_{100-x}Re_x$ alloys. At 2 K the pinning force density in this sample reaches a maximum of $2 \times 10^8$ N m-3. The peak pinning force density in the as-cast $Mo_{75}Re_{25}$ and $Mo_{60}Re_{40}$ alloys is only about one-fifth of this number. Due to this large pinning forces density in the as-cast $Mo_{52}Re_{48}$ alloy, fluxcompression in this alloy during the FCC experiments is likely to be much higher than our other alloys. The large local magnetic pressure (or vortex pressure) [43] generated due to this flux-compression leads to a highly metastable flux-line arrangement in this alloy. The reversal of temperature during the changing of the measurement protocol from FCC to FCW introduces energy fluctuations in the system, and this, we believe, tends to take the system towards the equilibrium fluxline arrangement. As a result flux-lines leak out of the strongpinning locations and there is a reduction of M when the temperature is increased from 2 K during the FCW experiments. It may be noted in this context that the decrease of M with increasing T during the FCW measurements was observed in the LFPME regime of several superconductors [1, 14, 19], and the phenomenon has been analyzed theoretically as well [1]. These theoretical explanations, however, do not seem to be suitable for the FCW M(T) behaviour in the HFPME regime of the as-cast $Mo_{52}Re_{48}$ alloy. Among the other materials exhibiting HFPME, the decrease of M with increasing T during the FCW measurements in the superconducting state of $Ba(Fe_{1-x}Co_x)_2As_2$ is related to the presence of Fe-ions in that material. We have already ruled out the possibility of the presence of magnetic impurity in our alloys. On the other hand, the decrease of M with increasing T during the FCW measurements has also been observed in the HFPME regime of the $YBa_2Cu_3O_{7-\delta}$ superconductors [8, 9, 25, 43], and there it appears to be related to the flux compressed state. In the absence of such highly metastable flux-compressed state, the FCW M(T) curve may be a weak function or an increasing function of T respectively-depending on whether the $T_{CR}$ is far lower than or close to the superconducting to normal phase transition.

**Summary and conclusions**

We have studied the temperature and field dependence of magnetization of a series of superconducting $Mo_{100-x}Re_x$ alloys, namely the as-cast samples of $Mo_{75}Re_{25}$, $Mo_{60}Re_{40}$, $Mo_{52}Re_{48}$ alloys and the annealed $Mo_{60}Re_{40}$ alloy, and have also performed complementary studies on the temperature dependence of the electrical resistivity and heat capacity of these alloys in different constant magnetic fields. In the temperature dependence of magnetization measured in different magnetic fields we have observed the following characteristic signatures of HFPME in these alloys: (i) the field cooled magnetization while warming up lies above the field cooled magnetization measured while cooling down, and (ii) field cooled magnetization increases continuously (exhibits relaxation effects) even at constant field and temperature. The linearity of the isothermal field dependence of magnetization of our alloys, measured just above the $T_C$, rules out the presence of magnetic impurities in the alloys. We have also shown experimentally that the HFPME in the $Mo_{100-x}Re_x$ alloys is not because of surface superconductivity, and this is in agreement with the existing literature which suggests that HFPME in general may not be related to the surface related effects [9, 30, 44]. Our results indicate that the HFPME in the $Mo_{100-x}Re_x$ alloys is related to the inhomogeneous distribution of strong and weak flux-line pinning centres, and the flux-compression resulting due to the same while cooling down the samples in the presence of high magnetic fields. Accordingly in the present as-cast $Mo_{75}Re_{25}$ and

$Mo_{60}Re_{40}$ alloys (single-phase), where the major pinning centres are more homogeneously distributed, the HFPME is observed in a comparatively lower field-temperature regime. The annealed $Mo_{60}Re_{40}$ and as-cast $Mo_{52}Re_{48}$ alloys (multiphase), on the other hand, have more inhomogeneous distribution of the stronger pinning centres, and accordingly they exhibit HFPME in higher field-temperature regime.

## Acknowledgments

We thank Shri R K Meena for his help in sample preparation, Dr Gurvinderjit Singh for the XRD measurements, and Dr R Rawat and Dr V K Sharma for doing some of the electrical resistivity measurements. We also thank Shri Rakesh Kaul and Dr M A Manekar for their help in optical metallography.